\documentclass[prd,notitlepage,superscriptaddress,nofootinbib,preprintnumbers]{revtex4-1}

\usepackage{amsmath}
\usepackage{amsfonts}
\usepackage{graphicx}
\usepackage[utf8]{inputenc}
\usepackage{slashed}

\usepackage{hyperref}
\hypersetup{colorlinks, linkcolor = [rgb]{0,0.0,0.75}, citecolor = [rgb]{0,0.0,0.75}, urlcolor = [rgb]{0,0.0,0.75}}

\makeatletter
\g@addto@macro\bfseries{\boldmath}
\makeatother

\DeclareMathOperator{\Tr}{Tr}
\DeclareMathOperator{\sgn}{sgn}
\DeclareMathOperator{\Ci}{Ci}
\newcommand{\MSbar}{\overline{\text{MS}}}
\newcommand{\cO}{\mathcal{O}}
\newcommand{\cL}{\mathcal{L}}

\begin{document}

\preprint{CERN-TH-2020-028}

\title{Improvement, generalization, and scheme conversion of Wilson-line operators on the lattice in the auxiliary field approach}

\author{Jeremy~R.~Green}
\email{jeremy.green@cern.ch}
\affiliation{Theoretical Physics Department, CERN, 1211 Geneva 23, Switzerland}
\author{Karl~Jansen}
\affiliation{NIC, Deutsches Elektronen-Synchrotron, 15738 Zeuthen, Germany}
\author{Fernanda~Steffens}
\affiliation{Institut für Strahlen- und Kernphysik, Universität Bonn, Nussallee 14--16, 53115 Bonn, Germany}

\date{\today}

\begin{abstract}
  Nonlocal quark bilinear operators connected by link paths are used
  for studying parton distribution functions (PDFs) and transverse
  momentum-dependent PDFs of hadrons using lattice QCD. The
  nonlocality makes it difficult to understand the renormalization and
  improvement of these operators using standard methods. In previous
  work, we showed that by introducing an auxiliary field on the
  lattice, one can understand an on-axis Wilson-line operator as the
  product of two local operators in an extended theory. In this paper,
  we provide details about the calculation in perturbation theory of
  the factor for conversion from our lattice-suitable renormalization
  scheme to the $\MSbar$ scheme. Extending our work, we study Symanzik
  improvement of the extended theory to understand the pattern of
  discretization effects linear in the lattice spacing, $a$, which are
  present even if the lattice fermion action exactly preserves chiral
  symmetry. This provides a prospect for an eventual $O(a)$
  improvement of lattice calculations of PDFs. We also generalize our
  approach to apply to Wilson lines along lattice diagonals and to
  piecewise-straight link paths.
\end{abstract}

\maketitle

\section{Introduction}

Calculating parton distribution functions (PDFs) of hadrons using
lattice QCD is challenging. The most direct definition using a bilocal
light cone operator is not accessible because lattice QCD is
formulated in Euclidean space. The traditional approach is to compute
Mellin moments of PDFs, which are obtained from matrix elements of
local twist-two operators, but higher moments are problematic because
of mixing with lower dimensional operators and increasing noise. In
recent years, there has been a renewed interest in alternative
approaches that use matrix elements of nonlocal operators that can be
related to PDFs via perturbatively computable factorization
formulas~\cite{Monahan:2018euv, Cichy:2018mum}. Of these, the most
widely studied has been quasi-PDFs, proposed in
Ref.~\cite{Ji:2013dva}, which use matrix elements of the nonlocal
equal-time operator
\begin{equation}
  \cO_\Gamma(x,\xi,n) \equiv \bar\psi(x+\xi n)\Gamma W(x+\xi n,x)\psi(x),
\end{equation}
where $\psi$ and $\bar\psi$ are spatially separated by distance $\xi$
in direction $n$ and connected by a straight Wilson line $W$. In
recent years two of us have been involved in studies of quasi-PDFs by
the Extended Twisted Mass Collaboration
(ETMC)~\cite{Alexandrou:2015rja, Alexandrou:2016jqi,
  Alexandrou:2017huk, Alexandrou:2018pbm, Alexandrou:2018eet,
  Alexandrou:2019lfo, Chai:2019rer, Alexandrou:2019dax}. Additional
studies of quasi-PDFs and other observables defined using $\cO_\Gamma$
are given in Refs.~\cite{Lin:2014zya, Chen:2016utp, Zhang:2017bzy,
  Chen:2017mzz, Orginos:2017kos, Lin:2017ani, Chen:2017lnm,
  Chen:2017gck, Chen:2018xof, Chen:2018fwa, Lin:2018qky,
  Karpie:2018zaz, Liu:2018hxv, Chen:2019lcm, Izubuchi:2019lyk,
  Joo:2019jct}.

Some of the difficulties in the quasi-PDF approach arise from the use
of a nonlocal operator. In Refs.~\cite{Green:2017xeu, Ji:2017oey}, the
auxiliary field approach~\cite{Craigie:1980qs, Dorn:1986dt} was used
to represent the nonlocal operator as the product of two local
operators in an extended theory\footnote{A similar analysis for
  gluonic Wilson-line operators in the continuum was done in
  Refs.~\cite{Dorn:1981wa, Dorn:1986dt, Wang:2017eel, Wang:2017eel,
    Zhang:2018diq}.}. This makes it possible to understand the
renormalization properties of $\cO_\Gamma$ using standard techniques
applied to the local operators. Specifically, the auxiliary field
$\zeta(\xi)$, which is defined only along the line $x+\xi n$, is given
the Lagrangian
\begin{equation}
  \cL_\zeta = \bar\zeta(n\cdot D + m)\zeta,
\end{equation}
and $\cO_\Gamma$ is obtained using the local operator
$\phi\equiv\bar\zeta\psi$:
\begin{equation}
  \cO_\Gamma(x,\xi,n) = \left\langle \bar\phi(x+\xi n)\Gamma\phi(x)\right\rangle_\zeta,
\end{equation}
for $m=0$ and $\xi>0$.

Following the lattice theory for a static quark~\cite{Eichten:1989zv,
  Sommer:2010ic}, in Ref.~\cite{Green:2017xeu} we also defined a
lattice action for the auxiliary field with $n=\pm\hat\mu$ pointing
along one of the lattice axes,
\begin{equation}
    S_\zeta^\text{lat} = a\sum_\xi \frac{1}{1+am_0} \bar\zeta(x+\xi n)[\nabla_n+m_0]\zeta(x+\xi n),\qquad
    \nabla_n \equiv \begin{cases}
      n\cdot \nabla^* = \nabla^*_\mu & \text{if }n=+\hat\mu,\\
      n\cdot \nabla  = -\nabla_\mu   & \text{if }n=-\hat\mu,
    \end{cases}
\end{equation}
where $\nabla$ and $\nabla^*$ are the forward and backward lattice
covariant derivatives, respectively, and $a$ is the lattice
spacing. This enabled us to identify that the operator mixing observed
in one-loop lattice perturbation theory~\cite{Constantinou:2017sej} is
caused by mixing between $\phi$ and the operator $\slashed{n}\phi$,
which is allowed when the lattice fermion action breaks chiral symmetry.

In addition, we presented in Ref.~\cite{Green:2017xeu} a scheme for
nonperturbative renormalization of $\cO_\Gamma$, called RI-xMOM, which
proceeds by renormalizing the auxiliary field action and the local
composite operator $\phi$. In Section~\ref{sec:matching}, we present
the calculation using perturbation theory of the scheme conversion
from RI-xMOM to $\MSbar$, the result of which was reported in
Ref.~\cite{Green:2017xeu}.

In Section~\ref{sec:improvement} we supplement our
previous work by applying the Symanzik improvement
program~\cite{Symanzik:1983dc} to analyze $O(a)$ lattice
artifacts. Finally, in Section~\ref{sec:general} we present
generalizations of the auxiliary field approach to operators with
gauge connections that are not straight lines and for off-axis gauge
connections. Conclusions are presented in Section~\ref{sec:conclusions}.
In Appendix~\ref{app:static} we relate our results on
improvement to previous work done for the static quark theory and in
Appendix~\ref{app:whole_op} we compare with the improvement program
based on the whole-operator approach for $\cO_\Gamma$.

\section{Scheme conversion}
\label{sec:matching}

In this section we summarize the approach for nonperturbative
renormalization that we introduced in Ref.~\cite{Green:2017xeu} and
provide details of the perturbative calculation of the conversion
factor to the $\MSbar$ scheme. We stress that the results in
Section~\ref{sec:improvement}, which deals with the improvement of
the auxiliary theory and of $\cO_\Gamma$, and Section~\ref{sec:general},
which generalizes our approach to a broader range of operators, do not
depend on the use of a particular renormalization scheme and that the
auxiliary field framework can be used quite broadly for understanding
Wilson-line operators.

The RI-xMOM scheme is based on a variant
of the Rome-Southampton method~\cite{Martinelli:1994ty} for
nonperturbative renormalization. The essential feature is the
definition of renormalization conditions that can be imposed both
nonperturbatively on the lattice and in dimensionally regularized
perturbation theory, which allows for conversion to the $\MSbar$
scheme. In Landau gauge, we make use of the position-space bare
$\zeta$ propagator,
\begin{equation}
  S_\zeta(\xi) \equiv \langle \zeta(x+\xi n)\bar\zeta(x)\rangle_{\text{QCD}+\zeta}
  = \langle W(x+\xi n,x)\rangle_\text{QCD},
\end{equation}
the momentum-space bare quark propagator,
\begin{equation}
  S_\psi(p) \equiv \int d^4x e^{-ip\cdot x} \langle \psi(x) \bar\psi(0) \rangle,
\end{equation}
and the mixed-space bare Green's function for $\psi$:
\begin{equation}
  G(\xi,p) \equiv \int d^4x e^{ip\cdot x} \langle \zeta(\xi n)\phi(0)\bar\psi(x) \rangle_{\text{QCD}+\zeta}.
\end{equation}
These renormalize as
\begin{align}
  S_\zeta^R(\xi) &= Z_\zeta e^{-m|\xi|} S_\zeta(\xi),\\
  S_\psi^R(p) &= Z_\psi S_\psi(p), \\
  G^R(\xi,p) &= Z_\phi\sqrt{Z_\zeta Z_\psi} e^{-m|\xi|} G(\xi,p).
\end{align}

For the quark field renormalization, we adopt the standard RI$'$-MOM
condition,
\begin{equation}
  \left.\frac{-i}{12 p^2 Z_\psi^\text{RI}} \Tr\left[ S_\psi^{-1}(p) \slashed{p} \right]\right|_{p^2=\mu^2} = 1.
\end{equation}
The remaining conditions are imposed at momentum $p_0$ and distance
$\xi_0$, which define a family of renormalization schemes at scale
$\mu^2=p_0^2$ that depend on the dimensionless quantities $y\equiv
|p_0|\xi_0$ and $z\equiv p_0\cdot n/|p_0|$:
\begin{gather}
 -\left.\frac{d}{d\xi} \log \Tr S_\zeta(\xi)\right|_{\xi=\xi_0} + m = 0,\\
 \frac{Z_\zeta^\text{RI}}{3} \frac{[\Tr S_\zeta(\xi_0)]^2}{\Tr S_\zeta(2\xi_0)} = 1,
 \label{eq:cond_zeta}\\
 \frac{1}{12} \frac{Z_\phi^\text{RI}}{\sqrt{Z_\zeta^\text{RI} Z_\psi^\text{RI}}}
    \Re\Tr\left[ S_\zeta^{-1}(\xi_0)G(\xi_0,p_0)S_\psi^{-1}(p_0) \right] = 1.
 \label{eq:cond_phi}
\end{gather}
Note that the last two conditions have been formulated to eliminate
dependence on the linearly divergent counterterm $m$.

\begin{figure}
  \centering
  \includegraphics[width=0.4\columnwidth]{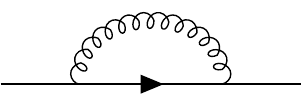}\hspace{0.1\columnwidth}
  \includegraphics[width=0.4\columnwidth]{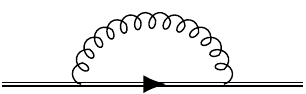}
  \caption{One-loop corrections to the quark (left) and
    auxiliary-field (right) propagators.}
  \label{fig:prop}
\end{figure}

For one-loop conversion to the $\MSbar$ scheme, we work in
$d$-dimensional Euclidean space with dimensional regularization and
use general covariant gauge with gauge parameter $\lambda$. At
one-loop order, $S_\zeta$ and $S_\psi$ are depicted in
Fig.~\ref{fig:prop}. We generically define conversion factors as
$C_X\equiv Z_X^{\MSbar}/Z_X^\text{RI}$. In the case of the quark
field, this has been computed e.g.\ in Ref.~\cite{Sturm:2009kb}:
$C_\psi=1-\frac{\alpha_sC_F}{4\pi}\lambda + O(\alpha_s^2)$, where
$C_F=4/3$ for QCD.

The free gluon propagator takes the form
\begin{equation}
  D^{AB}_{\mu\nu}(p) = \frac{\delta^{AB}}{p^2}\left[\delta_{\mu\nu}
 - (1-\lambda)\frac{p_\mu p_\nu}{p^2}\right].
\end{equation}
For $S_\zeta$, we will use it in position space, which is given in Ref.~\cite{Dorn:1986dt}:
\begin{equation}
  \int \frac{d^dp}{(2\pi)^d} e^{ip\cdot x} D^{AB}_{\mu\nu}(p)
= \delta^{AB}\frac{1+\lambda}{2}
\frac{\delta_{\mu\nu}\Gamma(\tfrac{d}{2}-1)}{4\pi^{d/2}(x^2)^{d/2-1}}
+ \delta^{AB}(1-\lambda)
\frac{x_\mu x_\nu \Gamma(\tfrac{d}{2})}{4\pi^{d/2}(x^2)^{d/2}}.
\end{equation}
Together with the tree-level $\zeta$ propagator,
\begin{equation}
  S_\zeta^\text{tree}(\xi) = \theta(\xi)e^{-m\xi},
\end{equation}
the loop integral for the auxiliary-field propagator is
straightforward. We obtain in $\MSbar$ at scale $\mu$:
\begin{equation}
  S_\zeta^{\MSbar}(\xi;\mu) = e^{-m\xi}\biggl( 1
+ \frac{\alpha_s C_F}{2\pi}\left[
      2 + (3-\lambda)\left(\gamma_E + \log\frac{\xi\mu}{2}\right)\right]
\biggr),
\end{equation}
where $\gamma_E$ is the Euler-Mascheroni constant.  This agrees with
the $O(\alpha_s)$ term in Ref.~\cite{Chetyrkin:2003vi}.
Eq.~\eqref{eq:cond_zeta} implies
\begin{equation}
  C_\zeta(y) = \frac{S_\zeta^{\MSbar}(\xi_0;\mu)^2}{S_\zeta^{\MSbar}(2\xi_0;\mu)}
 = 1 
+ \frac{\alpha_s C_F}{2\pi}\left[
      2 + (3-\lambda)\left(\gamma_E + \log\frac{y}{4}\right)\right]
+ O(\alpha_s^2).
\end{equation}

\begin{figure}
  \centering
  \includegraphics[width=0.3\columnwidth]{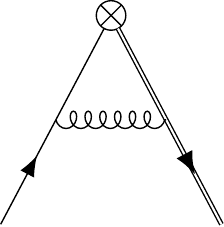}
  \caption{One-loop correction to the vertex function of the bilinear
    $\phi$.}
  \label{fig:vertex}
\end{figure}

For the one-loop vertex function (Fig.~\ref{fig:vertex}), we use the
free quark and gluon propagators in mixed space: position space
parallel to $n$ and momentum space for the $d-1$ orthogonal
dimensions. We decompose a general vector as $v\equiv v_nn + v_\perp$,
where $v_n=n\cdot v$. The gluon propagator takes the form
\begin{equation}
\bar D_{\mu\nu}^{AB}(x_n,p_\perp) \equiv
  \int \frac{dp_n}{2\pi} e^{ip_nx_n}D^{AB}_{\mu\nu}(p)
= \delta^{AB}\frac{e^{-|x_n||p_\perp|}}{2|p_\perp|}
  \biggl[\\
\delta_{\mu\nu} - \frac{1-\lambda}{2|p_\perp|^2}\left(
|x_n||p_\perp|p^*_\mu p^*_\nu + p_{\perp\mu}p_{\perp\nu} + |p_\perp|^2n_\mu n_\nu
\right)\biggr],
\end{equation}
where $p^*\equiv p_\perp+i\sgn(x_n)|p_\perp|n$. Similarly, for the quark
propagator we obtain
\begin{equation}
\bar S_\psi^\text{tree}(x_n,p_\perp)\equiv
\int \frac{dp_n}{2\pi} e^{ip_nx_n} \frac{-i\slashed{p}}{p^2}
 = \delta^{ab}e^{-|x_n||p_\perp|}\frac{-i\slashed{p}^*}{2|p_\perp|}.
\end{equation}
We write the quantity in Eq.~\eqref{eq:cond_phi}, as
\begin{equation}
  \Lambda_{\phi}(\xi,p) \equiv S_\zeta^{-1}(\xi) G(\xi,p)S_\psi^{-1}(p).
\end{equation}
This is similar to an amputated vertex function, except that the
$\zeta$-leg ``amputation'' is done in position space rather than the
usual momentum space. At one-loop order, this takes the form
\begin{multline}
  \Lambda_\phi(\xi,p) = 1 + (ig)^2\mu^{4-d} S_\zeta^\text{tree}(\xi)^{-1}
 \int_0^\xi d\xi_1 \int dx_n e^{ix_np_n}\int\frac{d^{d-1}k_\perp}{(2\pi)^{d-1}}\biggl[\\
S_\zeta^\text{tree}(\xi-\xi_1)T^An_\mu S_\zeta^\text{tree}(\xi_1) \bar S_\psi^\text{tree}(-x_n,p_\perp-k_\perp)T^B\gamma_\nu \bar D^{AB}_{\mu\nu}(\xi_1-x_n,k_\perp)\biggr].
\end{multline}
We simplify the calculation by restricting ourselves to the kinematics
$p_\perp=0$ and $p_n>0$ (i.e.\ $z=1$), which leads to
\begin{multline}
  \frac{1}{12}\Tr\Lambda_\phi(\xi,p) = 1 - g^2\mu^{4-d}C_F
 \int_0^\xi d\xi_1 \int dx_n e^{ix_np_n}\int\frac{d^{d-1}k_\perp}{(2\pi)^{d-1}}\biggl[
  e^{-|k_\perp|(|x_n|+|\xi_1-x_n|)}\\
\times \frac{-1}{4|k_\perp|}\bigl(\tfrac{1+\lambda}{2}\sgn(x_n)
+ \tfrac{1-\lambda}{2}|\xi_1-x_n||k_\perp|[\sgn(x_n)-\sgn(\xi_1-x_n)]\bigr)
\biggr].
\end{multline}
Eventually, we obtain
\begin{multline}
  \frac{1}{12}\Re\Tr\Lambda_{\phi}^{\MSbar}(\xi,p;\mu) = 1
+ \frac{\alpha_s C_F}{2\pi}\Biggl(
\lambda\left(1+\log\frac{\mu}{p_n}\right)
- 2\log 2 - \frac{1+\lambda}{2}\frac{\sin p_n\xi}{p_n\xi}
 - \frac{1-\lambda}{4}(\cos p_n\xi -3) \\
+ \left[(\lambda-2)\cos\frac{p_n\xi}{2}
  - \frac{1-\lambda}{4} p_n\xi\sin\frac{p_n\xi}{2}\right]\Ci\left(\frac{p_n\xi}{2}\right)
+ 2\Ci(p_n\xi) \Biggr),
\end{multline}
where $\Ci$ is the cosine integral function,
$\Ci(z) \equiv -\int_z^\infty \frac{\cos(t)}{t} dt$.

The $\phi$ conversion factor can be evaluated using
$C_\phi=\sqrt{C_\zeta
  C_\psi}\frac{1}{12}\Re\Tr\Lambda_{\phi}^{\MSbar}$ with the
appropriate kinematics. In Landau gauge ($\lambda=0$), we obtain
\begin{equation}
   C_\phi(y) = 1
+ \frac{\alpha_s C_F}{2\pi}\biggl[
  \frac{3}{2}\log\frac{y}{4}
  +\frac{3}{2}\gamma_E - 2\log 2 + \frac{7}{4} - \frac{\cos y}{4}
 \\ - \left(2\cos\frac{y}{2} - \frac{y}{4}\sin\frac{y}{2}\right)
\Ci\left(\frac{y}{2}\right) + 2\Ci(y)\biggr]
 + O(\alpha_s^2).
\end{equation}
We also find an anomalous dimension consistent with the leading term
obtained first in the auxiliary field
theory~\cite{Craigie:1980qs, Dorn:1986dt} and later also for the
static-light current~\cite{Shifman:1986sm, Politzer:1988wp,
  Politzer:1988bs, Chetyrkin:2003vi}.

To match the auxiliary field mass, one can use the higher-order
results from Ref.~\cite{Chetyrkin:2003vi}. Comparing conventions, our
$S_\zeta$ corresponds to their coordinate-space $i\tilde S_r$ and our
$Z_\zeta$ corresponds to their $\tilde Z_Q^{-1}$. That reference gives
\begin{equation}
  \log S_\zeta^{\MSbar}(\xi;\mu_\xi)=f_\zeta(\alpha_\xi),
\end{equation}
where $\mu_\xi\equiv \frac{2}{\xi}e^{-\gamma_E}$, $\alpha_\xi\equiv
\alpha_s(\mu_\xi)$, and $f_\zeta(x)$ is a series in $x$ given up to
$O(x^3)$. Together with the anomalous dimension $\tilde\gamma_Q\equiv
\frac{d\log\tilde Z_Q}{d\log\mu}$ and the beta function
$\beta\equiv\frac{-1}{2}\frac{d\log\alpha_s}{d\log\mu}$, we obtain
\begin{equation}
  -\frac{d}{d\xi}\log S_\zeta^{\MSbar}(\xi;\mu_0) = \frac{1}{\xi}\left[
    \tilde\gamma_Q(\alpha_\xi)
    - 2\alpha_\xi\beta(\alpha_\xi)f_\zeta'(\alpha_\xi)\right],
\end{equation}
which is independent of the scale $\mu_0$. Perturbatively, this is
limited by $\tilde\gamma_Q$, which has been computed to order
$\alpha_s^3$~\cite{Chetyrkin:2003vi,Melnikov:2000zc}.

\section{Improvement}
\label{sec:improvement}

In the Symanzik approach~\cite{Symanzik:1983dc, Luscher:1996sc}, the
lattice theory is described by a continuum effective theory that is an
expansion in powers of the lattice spacing, where each term has the
same symmetries as the lattice theory. The same is also done for
composite operators. The idea of \emph{improvement} is to add higher
dimensional operators in order to tune the parameters of the Symanzik
theory such that the leading [e.g.\ $O(a)$] term in the continuum
extrapolation of every correlation function is eliminated.

For the lattice fermion action we take Wilson twisted mass fermions,
working in the twisted basis with the fermion doublet $\chi$, although
we will also consider chiral fermions. The leading continuum
Lagrangian is twisted mass QCD,
\begin{equation}
  \cL_\chi^{(0)} = \bar\chi(\slashed{D} + m_q + i\mu_q\gamma_5\tau^3)\chi.
\end{equation}
At the next order $\cL_\chi^{(1)}$ contains terms that can be absorbed
into the parameters of the leading Lagrangian along with one
nontrivial one, the well-known ``clover''
term~\cite{Sheikholeslami:1985ij, Luscher:1996sc, Shindler:2007vp}
associated with chiral symmetry breaking.  For the auxiliary field in
the continuum, we have
\begin{equation}
  \cL_\zeta^{(0)} = \bar\zeta(n\cdot D + m)\zeta.
\end{equation}

Exact symmetries of the lattice theory include the little group of
hypercubic rotations that preserve $n$ and the $U(1)$ charge symmetry
for the auxiliary field. There are also several spurionic
symmetries~\cite{Dorn:1986dt, Shindler:2007vp, Chen:2017mie}:
\begin{enumerate}
\item General hypercubic rotations, together with the rotation of $n$.
\item Parity with respect to the axis $n$, i.e.\ the negation of the
  part of space-time vectors orthogonal to $n$, together with the
  negation of $\mu_q$.
\item Time reversal with respect to the axis $n$, together with the
  negation of $\mu_q$ and $n$.
\item Charge conjugation, together with the negation of
  $n$. Specifically, the auxiliary field transforms as
  \begin{equation}
    \zeta_n(x) \to \bar\zeta_{-n}(x)^T,\qquad
    \bar\zeta_n(x) \to \zeta_{-n}(x)^T.
  \end{equation}
\item Flavor $SU(2)$, together with a rotation of the twisted mass
  term $\mu_q\tau^3 \to
  \mu_qe^{i\alpha^a\tau^a}\tau^3e^{-i\alpha^a\tau^a}$.
\item For a chiral fermion action, $SU(2)$ axial transformations
  together with a rotation of the total mass term
  $m_q+i\mu_q\gamma_5\tau^3$.
\end{enumerate}

Following~\cite{Kurth:2000ki}, the next-order auxiliary field
Lagrangian has the form
\begin{equation}
  \cL_\zeta^{(1)} = a( b_\zeta m_q^2 + b'_\zeta \mu_q^2 )\bar\zeta\zeta,
\end{equation}
where $b_\zeta$ and $b'_\zeta$ are $O(g_0^4)$. These terms produce a
quark-mass dependence in the auxiliary field mass counterterm. This
effect could be significant if charm quarks are included in the
lattice action.

\subsection{Local bilinear operator}

To simplify the study of improvement, we consider a bilinear defined
in the twisted basis, $\phi=\bar\zeta\chi$. The operator
$\slashed{n}\phi$ transforms in the same way under all of the above
symmetries except for $SU(2)$ axial. Therefore at leading order the
renormalized operator is given by
\begin{equation}
  \phi_R = Z_\phi(\phi + r_\text{mix}\slashed{n}\phi),
\end{equation}
where $r_\text{mix}=0$ for a chiral action.

For the $O(a)$ contributions, we follow the study of the static-light
axial current in Ref.~\cite{Kurth:2000ki}. However, because of the
equation of motion $(\slashed{D}+m_q+i\mu_q\tau_3\gamma_5)\chi=0$,
there is some freedom in defining the on-shell improvement terms. In
particular, we could use a derivative orthogonal to $n$,
$\slashed{D}_\perp=(\delta_{\mu\nu}-n_\mu n_\nu)\gamma_\mu D_\nu$ as
in Ref.~\cite{Kurth:2000ki}. However, for the case of quasi-PDFs it
may be better to use a derivative along $n$: this would keep the
improved operator from extending across more than one time slice and
could potentially allow $O(a)$ improvement to be applied to existing
data. Furthermore, we use the equations of motion\footnote{Note that
  the improved operators will be defined in the case where the bare
  mass $m$ of the $\zeta$ field is set to zero. In the case where it
  is nonzero, the additional contribution here could be absorbed into
  $Z_\phi$.} for $\zeta$ to write $\bar\zeta n\cdot D\chi =
n\cdot\partial(\bar\zeta\chi)$. At $O(a)$, we obtain the improved
operator,
\begin{equation}
\begin{aligned}
  \phi_{R,I} &= Z_\phi(\phi + r_\text{mix}\slashed{n}\phi + a\phi_{1,m} + a\phi_{1,D}),\\
  \phi_{1,m} &= \left[ (b_\phi + \bar b_\phi\slashed{n})\slashed{n}m_q
    + (b'_\phi + \bar b'_\phi\slashed{n})\slashed{n}i\mu_q\tau^3\gamma_5
  \right]\phi,\\
  \phi_{1,D} &= (c_\phi + \bar c_\phi\slashed{n})n\cdot\partial \phi.
\end{aligned}
\end{equation}
For a chiral action, $\bar b_\phi$, $\bar b'_\phi$, and $\bar
c_\phi$ vanish and $b_\phi=b'_\phi$ but there still exist terms at
$O(a)$ that can not be excluded. Note that this expression assumes
only a single doublet of fermions is present; for additional
nondegenerate fermions there can be additional terms involving, for
example, the trace of the mass matrix, as discussed in
Ref.~\cite{Bhattacharya:2005rb}. For $\bar\phi=\bar\chi\zeta$, we
apply charge conjugation and then relabel $n\to-n$ to obtain
\begin{equation}
\begin{aligned}
  \bar\phi_{R,I} &= Z_\phi(\bar\phi + r_\text{mix}\bar\phi\slashed{n} + a\bar\phi_{1,m} + a\bar\phi_{1,D})\\
  \bar\phi_{1,m} &= \bar\phi\left[ 
    m_q\slashed{n}(b_\phi + \bar b_\phi\slashed{n})
    + i\mu_q\tau^3\gamma_5\slashed{n}(b'_\phi + \bar b'_\phi\slashed{n})
  \right],\\
  \bar\phi_{1,D} &= -n\cdot\partial\bar\phi(c_\phi + \bar c_\phi\slashed{n}).
\end{aligned}
\end{equation}

One further consideration is if the lattice gauge links used for the
Wilson line are obtained using an anisotropic smearing, which breaks
some of the hypercubic rotations and allows additional
contributions. This is used in calculations by ETMC, where (for $n$ in
a spatial direction) the gauge links are smeared only in spatial
directions and not in time~\cite{Alexandrou:2019lfo}. There is again
some freedom in defining the on-shell improvement due to the equations
of motion for $\chi$. If smearing is not performed in the $t$
direction, then one possible form for the additional term is $(c'_\phi
+ \bar c'_\phi \slashed{n}) \gamma_t\bar\zeta D_t\chi$. We will not
explicitly consider this anisotropic case below, but the
generalization is straightforward.

\subsection{Operator for quasi-PDFs}

Using the fermion doublet $\chi$, we consider the operator
\begin{equation}
  \cO_{\Gamma\tau}(\xi)
  = \bar\chi(\xi n)\Gamma\tau W(\xi n,0)\chi(0)
  = \left\langle \bar\phi(\xi n)\Gamma\tau\phi(0)\right\rangle_\zeta,
\end{equation}
inserted at zero momentum, where $\Gamma$ is a generic spin matrix and
$\tau$ is a generic flavor matrix. The improved, renormalized operator
is given (for $\xi>0$) by
\begin{equation}
\begin{gathered}
\begin{aligned}
  \cO^{R,I}_{\Gamma\tau}(\xi) &=
  \langle \bar\phi_{R,I}(\xi n)\Gamma\tau \phi_{R,I}(0) \rangle_\zeta\\
  &= Z_\phi^2 e^{-m\xi}\biggl(
  \bar\chi(\xi n)X W(\xi n,0)\chi(0)\\
  &\qquad\qquad\quad + a \left.\frac{\partial}{\partial\eta} \bigl[
    \bar\chi(\xi n) W(\xi n,\eta n)(1 + r_\text{mix}\slashed{n})\Gamma\tau
    (c_\phi + \bar c_\phi\slashed{n}) \chi(\eta n) \bigr] \right|_{\eta=0} \\
  &\qquad\qquad\quad - a \frac{\partial}{\partial\xi}\bigl[
  \bar\chi(\xi n) W(\xi n,0) (c_\phi + \bar c_\phi\slashed{n})\Gamma\tau
  (1 + r_\text{mix}\slashed{n})\chi(0) \bigr]
\biggr) + O(a^2),
\end{aligned}\\
\begin{aligned}
  X &=
  \Gamma\tau + r_\text{mix}\{\Gamma,\slashed{n}\}\tau + r_\text{mix}^2 \slashed{n}\Gamma\slashed{n}\tau\\
  &\quad + am_q\bigl( 2\bar b_\phi \Gamma\tau + (b_\phi + r_\text{mix}\bar b_\phi)\{\Gamma,\slashed{n}\}\tau + 2b_\phi r_\text{mix} \slashed{n}\Gamma\slashed{n}\tau \bigr)\\
  &\quad + ia\mu_q\bigl( b'_\phi[\Gamma\tau,\slashed{n}\gamma_5\tau^3]
  + \bar b'_\phi\{\Gamma\tau,\gamma_5\tau^3\}
  + r_\text{mix}(
  \bar b'_\phi[\slashed{n}\Gamma\slashed{n}\tau,\slashed{n}\gamma_5\tau^3]
  + b'_\phi\{\slashed{n}\Gamma\slashed{n}\tau,\gamma_5\tau^3\}) \bigr).
\end{aligned}
\end{gathered}
\end{equation}
Since the operator is inserted at zero momentum, we can add a total
derivative to finally obtain
\begin{equation}\label{eq:improved}
\begin{gathered}
  \cO^{R,I}_{\Gamma\tau}(\xi) = Z_\phi^2 e^{-m\xi}\left[
    \cO_X(\xi) - a\frac{\partial}{\partial\xi}\cO_{\Gamma_D\tau}(\xi)\right]
  + O(a^2) + \text{total derivative},\\
  \Gamma_D = 2c_\phi\Gamma + (\bar c_\phi + c_\phi r_\text{mix})\{\Gamma,\slashed{n}\} + 2c_\phi r_\text{mix} \slashed{n}\Gamma\slashed{n}.
\end{gathered}
\end{equation}

For specific choices of $\Gamma$ and $\tau$, this expression
simplifies. Calculations of quasi-PDFs are typically done with a
flavor diagonal operator where $\tau$ commutes with $\tau^3$. It has
become standard to compute unpolarized quasi-PDFs using
$\Gamma=\gamma_\nu$ with $\nu$ satisfying $n_\nu=0$ and helicity
quasi-PDFs using $\Gamma=\slashed{n}\gamma_5$. In both of these cases,
$\slashed{n}\Gamma\slashed{n}=-\Gamma$ and all of the
(anti)commutators vanish, so that the improved operator becomes
\begin{equation}\label{eq:no_mixing}
\text{(no-mixing case)}\quad  \cO^{R,I}_{\Gamma\tau}(\xi) = Z_\phi^2 e^{-m\xi} \left[
  1 - r_\text{mix}^2 + 2am_q(\bar b_\phi - b_\phi r_\text{mix})
  - 2(c_\phi - \bar c_\phi r_\text{mix}) a\frac{\partial}{\partial\xi}\right]
\cO_{\Gamma\tau}(\xi),
\end{equation}
i.e.\ only the derivative operator (and no other mixing) contributes
at $O(a)$ and the only nontrivial effect of chiral symmetry breaking
is the dependence on $am_q$. Transversity quasi-PDFs are computed
using an operator that has $\Gamma=\slashed{n}\gamma_\nu$ in the
physical basis. If $\mu_q=0$, then Eq.~\eqref{eq:no_mixing} again
applies. On the other hand, when using twisted mass fermions at
maximal twist we must consider the corresponding twisted-basis
operator. For the case of a flavor diagonal transversity operator, the
twisted-basis operator has $\Gamma=i\slashed{n}\gamma_\nu\gamma_5$ and
suffers from equal-dimensional mixing because
$\{\Gamma,\slashed{n}\}=2i\gamma_\nu\gamma_5$. Using a flavor-changing
transversity operator would be advantageous: it appears the same in
the physical and twisted bases and the form of the improved operator
is given by Eq.~\eqref{eq:no_mixing}.

\subsubsection{Maximal twist}

In calculations done using Wilson twisted mass fermions, many
correlators benefit from automatic $O(a)$
improvement~\cite{Frezzotti:2003ni, Shindler:2007vp}. This means that
when tuned to maximal twist ($m_q=0$), the $O(a)$ contributions to
those correlation functions vanish and there is no need to explicitly
tune the improvement coefficients. For the case considered here, the
arguments behind automatic improvement do not eliminate all of the
$O(a)$ contributions, but they do eliminate the contributions that
vanish for chiral fermion actions.

We start by working at $O(a^0)$ and examining the equal-dimensional
mixing. To simplify the expressions, assume
$\slashed{n}\Gamma\slashed{n}=G_\Gamma\Gamma$, where $G_\Gamma=\pm
1$. We then write
\begin{align}
  \cO_{\Gamma\tau}^R(\xi) &= Z_\phi^2 e^{-m\xi}\left[
    (1 + G_\Gamma r_\text{mix}^2)\cO_{\Gamma\tau}(\xi)
    + r_\text{mix}\cO_{\{\Gamma,\slashed{n}\}\tau}(\xi)
    \right] + O(a),\\
  \cO_{\{\Gamma,\slashed{n}\}\tau}^{R}(\xi) &= Z_\phi^2 e^{-m\xi}\left[
    (1 + r_\text{mix}^2)\cO_{\{\Gamma,\slashed{n}\}\tau}(\xi)
    + 2(1 + G_\Gamma)r_\text{mix}\cO_{\Gamma\tau}(\xi)
    \right] + O(a).
\end{align}
Using these two equations we can eliminate
$\cO_{\{\Gamma,\slashed{n}\}\tau}$ and obtain
\begin{equation}\label{eq:autoimp}
  \cO_{\Gamma\tau}(\xi) = \frac{e^{m\xi}}{\tilde Z^2_{\phi,G_\Gamma}} \left[
    \cO_{\Gamma\tau}^{R}(\xi) - \frac{r_\text{mix}}{1+r_\text{mix}^2}
    \cO_{\{\Gamma,\slashed{n}\}\tau}^{R}(\xi) \right]
  + O(a),
\end{equation}
where $\tilde Z^2_{\phi,G}=Z_\phi^2 (1-r_\text{mix}^2)
(1-Gr_\text{mix}^2) / (1+r_\text{mix}^2)$.

We now consider the transformations
\begin{equation}
  \mathcal{R}_5^{1,2}:\begin{cases}
    \chi \to i\gamma_5 \tau^{1,2}\chi\\
    \bar\chi \to \bar\chi i\gamma_5 \tau^{1,2}
    \end{cases},
\end{equation}
which are chiral symmetries of continuum twisted mass QCD at maximal
twist. Clearly, if $\cO_{\Gamma\tau} \to R \cO_{\Gamma\tau}$ ($R=\pm
1$) under one of these transformations, then
$\cO_{\{\Gamma,\slashed{n}\}\tau} \to -R
\cO_{\{\Gamma,\slashed{n}\}\tau}$. Now, consider a correlation
function involving some hadronic interpolators and $\cO_{\Gamma\tau}$
that is invariant under $\mathcal{R}_5^{1,2}$. In the continuum, the
corresponding correlation function with
$\cO_{\{\Gamma,\slashed{n}\}\tau}$ will vanish because it is odd under
$\mathcal{R}_5^{1,2}$. On the lattice, for the renormalized operator
$\cO_{\{\Gamma,\slashed{n}\}\tau}^{R}$ the result must therefore be
$O(a)$ and this term can be neglected in
Eq.~\eqref{eq:autoimp}. Effectively, one can use
$\cO_{\Gamma\tau}^R(\xi) = \tilde
Z^2_{\phi,G_\Gamma}e^{-m\xi}\cO_{\Gamma\tau}(\xi) + O(a)$. This
justifies the calculation in Refs.~\cite{Alexandrou:2018eet,
  Alexandrou:2019lfo}, where equal-dimensional mixing for the
transversity operator was not explicitly treated.

Now, we move on to the $O(a)$ contributions. Specifically, we take the
correlation function of the product of $\cO_\Gamma$ inserted at zero
momentum with a renormalized $O(a)$-improved multilocal field
$\Phi$. The Symanzik expansion of the lattice correlator is given by
\begin{equation}
\begin{aligned}
  \bigl\langle \Phi \cO_{\Gamma\tau}(\xi) \bigr\rangle &=
  \frac{e^{m\xi}}{\tilde Z^2_{\phi,G_\Gamma}} \left\langle \Phi \left[ \cO_{\Gamma\tau}^R(\xi) - \frac{r_\text{mix}}{1+r_\text{mix}^2} \cO_{\{\Gamma,\slashed{n}\}\tau}^R(\xi) \right] \right\rangle_0 \\
  &\quad
  - \frac{e^{m\xi}}{\tilde Z^2_{\phi,G_\Gamma}} a \left\langle \Phi \left[ \cO_{\Gamma\tau}^R(\xi) - \frac{r_\text{mix}}{1+r_\text{mix}^2} \cO_{\{\Gamma,\slashed{n}\}\tau}^R(\xi) \right] \int d^4x \cL_\chi^{(1)}(x) \right\rangle_0 \\
  &\quad
  + a\left\langle \Phi \cO_{\Gamma\tau}^{(1)}(\xi) \right\rangle_0
  + O(a^2),
\end{aligned}
\end{equation}
where $\langle\cdots\rangle_0$ is evaluated in the continuum theory
and $\cO_{\Gamma\tau}^{(1)}(\xi)$ contains the $O(a)$ terms that
appear in Eq.~\eqref{eq:improved}. Assuming $\Phi\cO_{\Gamma\tau}$ is
invariant under $\mathcal{R}_5^{1,2}$ and $m_q=0$, many of the terms
vanish and we get
\begin{equation}\label{eq:symanzik_auto}
\begin{aligned}
\tilde Z^2_{\phi,G_\Gamma}e^{-m\xi} \bigl\langle \Phi \cO_{\Gamma\tau}(\xi) \bigr\rangle &=
  \left\langle \Phi \cO_{\Gamma\tau}^R(\xi) \right\rangle_0
  + \frac{r_\text{mix}}{1+r_\text{mix}^2}  a \left\langle \Phi \cO_{\{\Gamma,\slashed{n}\}\tau}^R(\xi) \int d^4x \cL_\chi^{(1)}(x) \right\rangle_0 \\
  &\quad
  + a\left\langle \Phi\left[
      c_{1,G_\Gamma} i\mu_q \cO^R_{[\Gamma\tau,\slashed{n}\gamma_5\tau^3]}(\xi)
      + c_{2,G_\Gamma} \frac{\partial}{\partial\xi} \cO^R_{\Gamma\tau}(\xi) \right] \right\rangle_0,
\end{aligned}
\end{equation}
for some prefactors $c_{1,\pm}$ and $c_{2,\pm}$. If the fermion action
is $O(a)$ improved by including a clover term\footnote{Note that the
  more recent twisted mass lattice ensembles generated by ETMC,
  including those at the physical pion mass, do include a clover
  term~\cite{Becirevic:2006ii, Abdel-Rehim:2015pwa,
    Alexandrou:2018egz}.}, then $\cL^{(1)}_\chi$ vanishes and one
can effectively obtain $O(a)$ improvement by using
\begin{equation}\label{eq:cO_autoimp}
\begin{gathered}
  \cO_{\Gamma\tau}^{R,I}(\xi) = \tilde Z^2_{\phi,G_\Gamma}e^{-m\xi}\left[
    \cO_{\tilde X}(\xi)
    - \tilde c_{\phi,G_\Gamma} a\frac{\partial}{\partial\xi}\cO_{\Gamma\tau}(\xi)
    \right],\\
    \tilde X = \Gamma\tau - i\tilde b'_{\phi,G_\Gamma} a\mu_q [\Gamma\tau,\slashed{n}\gamma_5\tau^3],
\end{gathered}
\end{equation}
where six parameters remain: $\tilde Z_{\phi,+}$, $\tilde
b'_{\phi,+}$, $\tilde c_{\phi,+}$, $\tilde Z_{\phi,-}$, $\tilde
b'_{\phi,-}$, and $\tilde c_{\phi,-}$. Usually the operator will have
a definite $G_\Gamma$, so that only half of the parameters can
contribute, and for many operators the commutator vanishes so that the
term proportional to $a\mu_q$ does not contribute. On the other hand,
if the fermion action is not $O(a)$ improved, then the second term in
Eq.~\eqref{eq:symanzik_auto} can be eliminated either by explicitly
treating the $O(a^0)$ mixing or by choosing an operator such that
$\{\Gamma,\slashed{n}\}=0$.

The term involving the $O(a)$ part of the fermion action survives in
Eq.~\eqref{eq:symanzik_auto} because the connection between
dimensional counting and breaking of chiral symmetry, which underlies
the usual arguments for automatic improvement, is broken by the
equal-dimensional mixing with $\cO_{\{\Gamma,\slashed{n}\}\tau}$. This
leads to another potential worry: if $\Phi$ is not $O(a)$ improved, as
is typically the case for interpolating operators, then an additional
nonvanishing term of the form $a\langle \Phi^{(1)}
\cO^R_{\{\Gamma,\slashed{n}\}\tau} \rangle_0$ can appear in
Eq.~\eqref{eq:symanzik_auto}. However, although this is an $O(a)$
contribution in the correlation function, it can be argued that it
will not contribute to the ground-state hadronic matrix element
determined at large Euclidean time separations, since the
latter is independent of the interpolating operator. To see this,
consider a correlation function using local hadron interpolators that
have been explicitly $O(a)$ improved. In this case, the Symanzik
expansion is given by Eq.~\eqref{eq:symanzik_auto} and the $O(a)$
terms can be eliminated by including a clover term in the fermion
action and tuning the improvement parameters in
Eq.~\eqref{eq:cO_autoimp}. Once the correlation function is free of
$O(a)$ effects, then, following the arguments in Section 3.2 of
Ref.~\cite{Frezzotti:2003ni}, we find that the matrix element obtained
at large time separations will be free of $O(a)$ effects. As the same
ground-state matrix element can be obtained using any interpolating
operator, the $O(a)$-improved ground-state matrix element can thus be
obtained from the large-time-separation limit of a correlation
function even if the interpolator is not $O(a)$ improved.

\subsection{Determining improvement coefficients}

In principle, parameters associated with breaking of chiral symmetry
($r_\text{mix}$, $\bar b_\phi$, $\bar b'_\phi$, $b_\phi-b'_\phi$, and
$\bar c_\phi$) can be determined using improvement conditions derived
from chiral Ward identities~\cite{Luscher:1996sc}; this approach was
used for nonperturbative determinations of closely-related parameters
in the static quark theory in Refs.~\cite{Aoki:1999ij,
  Hashimoto:2002pe, Palombi:2007dt}. For the remaining improvement
coefficients $b_\phi$ and $c_\phi$, the situation is more difficult as
there is no simple continuum physics condition to match onto: one
would have to numerically study the continuum extrapolation of
suitably chosen observables and tune the parameters to eliminate the
linear dependence on $a$. Alternatively, the parameters could be
computed in lattice perturbation theory, which has been done for the
static quark theory using a few different lattice
actions~\cite{Morningstar:1997ep, Ishikawa:1999yi, Kurth:2000ki,
  DellaMorte:2005nwx, Palombi:2007dt, Grimbach:2008uy,
  Ishikawa:2011dd}\footnote{The parameter $r_\text{mix}$ can also be
  determined from the perturbative study of $\cO_\Gamma(\xi)$ in
  Ref.~\cite{Constantinou:2017sej}. For the gauge actions common to
  both calculations, that reference agrees with
  Ref.~\cite{Ishikawa:1999yi}.}.

\section{General link paths}
\label{sec:general}

In this section we generalize our use of the auxiliary field approach
on the lattice to describe paths that have corners and paths that are
not along a lattice axis. In part, this will serve to understand under
which circumstances the assumption made in Refs.~\cite{Hagler:2009mb,
  Musch:2010ka, Musch:2011er, Engelhardt:2015xja} (and studied
empirically in Ref.~\cite{Yoon:2017qzo}), that the continuum
renormalization pattern~\cite{Dorn:1986dt} applies to lattice
calculations, is valid.

\subsection{Piecewise straight paths}

Nonlocal operators with link paths that are not straight are also used
for hadron structure; in particular, staple-shaped gauge connections
have been used for studying transverse momentum-dependent (TMD)
PDFs. If the path is made from a finite number of segments, each of
which is a Wilson line propagating along a lattice axis, then it is
straightforward to accommodate the nonlocal operator in the lattice
auxiliary field framework. An auxiliary field $\zeta_n$ must be
introduced for each segment, where $n$ is the corresponding
direction. In addition to suitable bilinears $\phi_{n_i}\equiv
\bar\zeta_{n_i}\psi$ and $\bar\phi_{n_f}$ for the endpoints, one must
also introduce cusp operators~\cite{Polyakov:1980ca, Craigie:1980qs}
$C_{n',n}=\bar\zeta_{n'}\zeta_n$ for each transition between segments.

\begin{figure}
  \centering
  \includegraphics[width=0.2\textwidth]{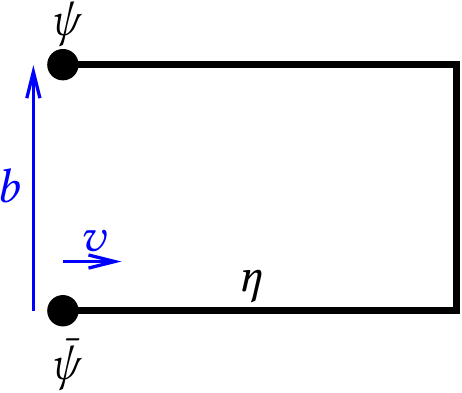}
  \caption{Staple-shaped operator.}
  \label{fig:staple}
\end{figure}

To be concrete, consider the TMD operator with quark fields separated
by the four-vector $b$ and the staple of height $\eta$ in the
direction of the unit vector $v$ that is orthogonal to $b$ (see
Fig.~\ref{fig:staple}):
\begin{equation}
  \cO^\text{TMD}_\Gamma = \bar\psi(0)\Gamma W(0,\eta v) W(\eta v,\eta v+b)
  W(\eta v+b,b) \psi(b).
\end{equation}
We introduce the auxiliary fields $\zeta_v$, $\zeta_{-v}$, and
$\zeta_{-\hat b}$, and obtain
\begin{equation}
  \cO^\text{TMD}_\Gamma = \left\langle \bar\phi_{-v}(0) \Gamma
   C_{-v,-\hat b}(\eta v) C_{-\hat b,v}(\eta v+b) \phi_v(b) \right\rangle_\zeta.
\end{equation}
In addition to renormalizing the action for the auxiliary fields and
the bilinears $\phi$, the cusp operators must also be renormalized.
The latter are not expected to mix. However, since the operators
$\phi_v$ and $\bar\phi_{-v}$ connect to auxiliary fields propagating
in opposite directions, the mixing pattern allowed by chiral symmetry
breaking is different from the straight-line operators used for
quasi-PDFs: $\cO^\text{TMD}_\Gamma$ can mix with
$\cO^\text{TMD}_{[\Gamma,\slashed{v}]}$. This was pointed out by one
of us in Ref.~\cite{Green_LatticePDF} and later confirmed at one-loop
order in lattice perturbation theory~\cite{Constantinou:2019vyb}. The
latter calculation also found that more generally, the pattern of
mixing depends only on the direction with which the Wilson lines hit
the quark fields, which is consistent with the prediction of the
auxiliary field approach. We note that a numerical study of
nonperturbative renormalization of staple-shaped operators was
recently done in Ref.~\cite{Shanahan:2019zcq}.

A RI-xMOM renormalization condition for cusp operators is
straightforward to formulate. This requires the position-space Green's
function for the cusp operator,
\begin{equation}
  G_C(\xi',\xi) \equiv \langle \zeta_{n'}(\xi' n') C_{n',n}(0) \bar\zeta(-\xi n) \rangle_{\text{QCD}+\zeta}
 = \langle W(\xi' n',0) W(0,-\xi n) \rangle_\text{QCD}.
\end{equation}
Performing a position-space ``amputation'', a possible condition is
\begin{equation}
 \frac{1}{3}\frac{Z_C}{Z_\zeta}\Tr\left[  S_{\zeta_{n'}}^{-1}(\xi') G_C(\xi',\xi) S_{\zeta_n}^{-1}(\xi) \right] = 1,
\end{equation}
when $\xi'=\xi=\mu^{-1}$. 

\subsection{Off-axis paths}

\begin{figure}
  \centering
  \includegraphics[width=0.25\textwidth]{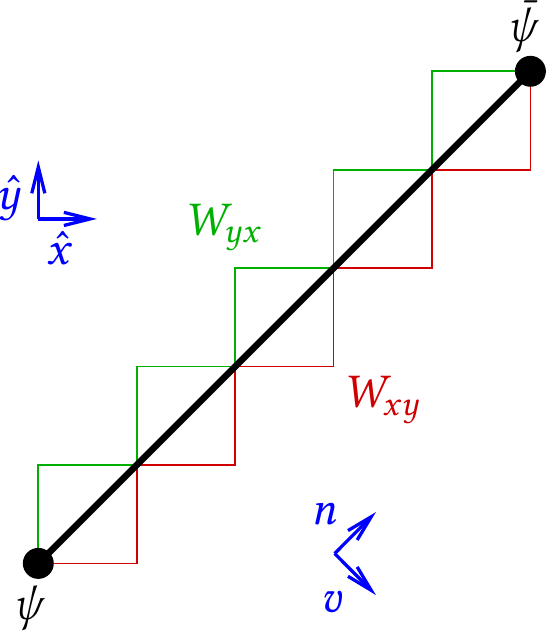}
  \caption{Off-axis operator with two different discretizations of the
    Wilson line.}
  \label{fig:offaxis}
\end{figure}

It is possible to somewhat relax the constraint that $n$ points along
a lattice axis. For example, let $n=\frac{1}{\sqrt{2}}(\hat x + \hat
y)$. On the lattice, one definition of the straight-line operator
is~\cite{Musch:2010ka}
\begin{equation}
  \cO_\Gamma^\text{lat}(0,\xi,n) \equiv \bar\psi(\xi n)\Gamma\frac{1}{2}\left[
    W_{xy}(\xi n,0) + W_{yx}(\xi n,0) \right] \psi(0),
\end{equation}
where $W_{xy}$ is formed from the zigzag product of gauge links
alternating between the $x$ and $y$ directions and vice-versa for
$W_{yx}$ (see Fig.~\ref{fig:offaxis}). This average is necessary so
that the operator has a simple transformation under $n$-parity. These
zigzag Wilson lines can be obtained as propagators of auxiliary fields
$\zeta_{xy}$ and $\zeta_{yx}$ that are defined using two-link
covariant derivatives along the $n$ direction, e.g.:
\begin{equation}
  \nabla_{xy}f(x) = \frac{1}{\sqrt{2}a} \left[
    f(x) - U_y^\dagger(x-a\hat y) U_x^\dagger(x-a(\hat x+\hat y))f(x-a(\hat x +\hat y)) \right].
\end{equation}
Because of rotational symmetry breaking on the lattice, the mass
counterterm $m_{xy}=m_{yx}$ will in general be different from the
on-axis case\footnote{In the continuum, the case of an auxiliary field
  propagating along a multi-cusp curve that approximates a smooth one
  was considered in Ref.~\cite{Craigie:1980qs}. In that case, when the
  number of cusps goes to infinity the action is equal to that of a
  field propagating along the smooth curve with an added mass term
  that accounts for the effect of the cusps. This is analogous to the
  situation here, where a straight line is approximated by zigzags
  that become infinitely many as the lattice spacing goes to
  zero.}. The quark bilinear takes the form
\begin{equation}
  \cO_\Gamma^\text{lat}(0,\xi,n) = \left\langle \frac{1}{2}\left[
      \bar\phi_{xy}(\xi n)\Gamma\phi_{xy}(0)
    + \bar\phi_{yx}(\xi n)\Gamma\phi_{yx}(0) \right] \right\rangle_\zeta,
\end{equation}
where $\phi_{xy}=\bar\zeta_{xy}\psi$, etc. Defining $\varphi_\pm
\equiv \frac{1}{\sqrt{2}}(\phi_{xy}\pm\phi_{yx})$, this can be
rewritten as
\begin{equation}
\cO_\Gamma^\text{lat}(0,\xi,n) 
= \left\langle \bar\varphi_\pm(\xi n)\Gamma \varphi_\pm(0) \right\rangle_\zeta,
\end{equation}
where the additional cross terms involving e.g.\ $\langle
\zeta_{xy}\bar\zeta_{yx}\rangle$ vanish. Under $n$-parity,
$\varphi_\pm \to \pm \slashed{n} \varphi_\pm$.

We find that additional mixing can occur. Defining
$v=\frac{1}{\sqrt{2}}(\hat x - \hat y)$, $\varphi_+$ mixes with
$\slashed{n}\varphi_+$, $\slashed{v}\varphi_-$, and
$\slashed{n}\slashed{v}\varphi_-$, the last of which is a chiral-even
mixing. This leads to correlation functions of the form $\langle
\bar\varphi_+ \varphi_- \rangle$, which contain a difference of Wilson
lines $W_{xy}-W_{yx}$. One expects that this difference will vanish in
the continuum limit. This expectation can be justified in the
auxiliary field approach: in the continuum, there is an $SU(2)$ flavor
symmetry relating $\zeta_{xy}$ and $\zeta_{yx}$. It is broken in the
next order of the Symanzik expansion, where a term of the form
$a\bar\zeta_{xy}G_{\mu\nu}n_\mu v_\nu\zeta_{xy}
-a\bar\zeta_{yx}G_{\mu\nu}n_\mu v_\nu\zeta_{yx}$ can appear in the
Lagrangian; this implies that 
$\langle\bar\varphi_+\varphi_-\rangle$ is $O(a)$. Therefore, even
though the mixing between $\varphi_+$ and operators containing
$\varphi_-$ is equal-dimensional, the most serious new mixing effect
in correlation functions is suppressed by at least $O(a)$. The
remaining effect is that even when using a chiral fermion action,
renormalization will be different for operators where
$\slashed{v}\slashed{n}\Gamma\slashed{n}\slashed{v}$ is equal to
$+\Gamma$ and those where it equals $-\Gamma$.

An alternative definition is to average the paths locally. Taking the
average of the two local link paths,
\begin{equation}
  U_\text{avg}(x) \equiv \frac{1}{2}\left[
    U_x(x)U_y(x+a\hat x) + U_y(x)U_x(a+a\hat y) \right],
\end{equation}
the lattice action for $\zeta_\text{avg}$ is defined using the
covariant derivative
\begin{equation}
  \nabla_\text{avg} f(x) = \frac{1}{\sqrt{2}a} \left[ f(x)
    - U_\text{avg}^\dagger(x-a(\hat x+\hat y))f(x-a(\hat x + \hat y)) \right].
\end{equation}
Again, the mass counterterm will in general be different from the
previous case. Then, using the bilinear
$\phi_\text{avg}=\bar\zeta_\text{avg}\psi$, the bilocal operator
$\langle \bar\phi_\text{avg}(\xi n)\Gamma
\phi_\text{avg}(0)\rangle_\zeta$ effectively averages over $2^N$ link
paths, where $\xi=\sqrt{2}Na$. In this case, the pattern of
equal-dimensional mixing is the same as the on-axis case.

\section{Conclusions}
\label{sec:conclusions}

The auxiliary field approach is an invaluable tool for understanding
the renormalization and improvement of Wilson-line operators on the
lattice. Generically, these nonlocal operators can be represented
using local operators in an extended theory involving auxiliary
fields. We have shown how this can be done for a variety of operators,
beyond the simplest ones involving straight on-axis link paths used
for quasi-PDF studies.

Using the Symanzik expansion, we were able to study discretization
effects and the form of the improved operators. As foreseen in
Ref.~\cite{Green:2017xeu}, we find that the leading effects are linear
in $a$ even if chiral symmetry is preserved on the lattice. Likewise,
we find that working at maximal twist does not automatically eliminate
the effects linear in $a$, but it can remove some of the contributions
and it can produce some simplification by reducing the number of
improvement coefficients. Our analysis provides a general framework
for the $O(a)$ improvement of nonlocal operators used in the quasi-PDF
approach for computing PDFs on the lattice.

In order to apply this improvement to a lattice calculation of
quasi-PDFs, the relevant coefficients for the choice of lattice action
must be determined. For some actions, many of these are already known from
lattice perturbation theory calculations done for the static quark
theory. There are plans to determine the coefficients for actions used
by ETMC~\cite{Constantinou_privcomm}, and we intend to study the
effect of improvement on the approach to the continuum limit. It will
be important to establish that discretization effects in quasi-PDF
calculations can be controlled.

\begin{acknowledgments}
F.\ S.\ was funded by DFG Project No.\ 392578569.
\end{acknowledgments}

\appendix

\section{Relation to static quark theory}
\label{app:static}

The static quark theory~\cite{Eichten:1989zv, Sommer:2010ic} is
defined using a spinor $Q$ that satisfies $Q=P_+Q$, where
$P_\pm=(1\pm\gamma_0)/2$. Its two spin degrees of freedom do not
couple in its action, and its propagator is the same as the
$\zeta$-field propagator with $n=\hat t$, multiplied by $P_+$. This
means that, after accounting for the projector, the static quark and
auxiliary field theories can be identified with each other. It will be
convenient to define the projected bilinears $\phi^\pm =
\frac{1}{2}(1\pm\slashed{n})\phi$, where the projectors become $P_\pm$
when $n=\hat t$. In our approach, neglecting a twisted mass term, the
renormalized $O(a)$-improved operators take the form
\begin{equation}\label{eq:imp_phipm}
  \phi^\pm_{R,I} = Z_\phi^\pm \left(
    1 \pm b_\phi^\pm am_q + c_\phi^\pm an\cdot\partial
    \right) \phi^\pm,
\end{equation}
where $Z_\phi^\pm=Z_\phi(1\pm r_\text{mix})$, $b_\phi^\pm=(b_\phi\pm
\bar b_\phi)/(1\pm r_\text{mix})$, and $c_\phi^\pm=(c_\phi\pm\bar
c_\phi)/(1\pm r_\text{mix})$.

In Ref.~\cite{Kurth:2000ki}, the time components of the static-light
axial and vector currents are defined as
\begin{align}
  A_0^\text{stat} &\equiv \bar\psi\gamma_0\gamma_5 Q = -\bar\psi P_- \gamma_5 Q,\\
  V_0^\text{stat} &\equiv \bar\psi\gamma_0 Q = \bar\psi P_+ Q.
\end{align}
By identifying which projector is contracted with $\bar\psi$, this
lets us make the identifications
\begin{equation}
  A_0^\text{stat}\to -\bar\phi^-\gamma_5,\qquad V_0^\text{stat}\to\bar\phi^+,
\end{equation}
and the corresponding renormalization factors can be equated:
$Z_A^\text{stat}=Z_\phi^-$ and $Z_V^\text{stat}=Z_\phi^+$. For the
static-light axial current, Ref.~\cite{Kurth:2000ki} identifies four
possible $O(a)$ improvement operators:
\begin{align}
(\delta A_0^\text{stat})_1 &\equiv \bar\psi \overset{\leftarrow}{D}_j\gamma_j\gamma_5 Q = \bar\psi \overset{\leftarrow}{\slashed{D}}_\perp P_- \gamma_5 Q,\\
(\delta A_0^\text{stat})_2 &\equiv \bar\psi \gamma_5 D_0 Q = \bar\psi P_- \gamma_5 D_0 Q,\\
(\delta A_0^\text{stat})_3 &\equiv \bar\psi \overset{\leftarrow}{D}_0 \gamma_5 Q = -\bar\psi \overset{\leftarrow}{D}_0 \gamma_0 P_- \gamma_5 Q,\\
(\delta A_0^\text{stat})_4 &\equiv m_q \bar\psi \gamma_0 \gamma_5 Q = -m_q \bar\psi P_- \gamma_5 Q.
\end{align}
At leading order, the equations of motion for $Q$ give $(\delta
A_0^\text{stat})_2=0$ and those for $\psi$ give $(\delta
A_0^\text{stat})_1-(\delta A_0^\text{stat})_3+(\delta
A_0^\text{stat})_4=0$; these are used to eliminate $(\delta
A_0^\text{stat})_{2,3}$. The improved, renormalized operator is then
given by
\begin{equation}\label{eq:imp_statA}
  (A_R^\text{stat})_0 = Z_A^\text{stat}(1 + b_A^\text{stat} am_q)(A_I^\text{stat})_0,
\quad (A_I^\text{stat})_0 = A_0^\text{stat} + a c_A^\text{stat} (\delta A_0^\text{stat})_1.
\end{equation}
We can make the identification
\begin{equation}
  (\delta A_0^\text{stat})_2 + (\delta A_0^\text{stat})_3 = \partial_0\left(
    \bar\psi P_- \gamma_5 Q \right) \to n\cdot\partial\bar\phi^-\gamma_5.
\end{equation}
To relate Eq.~\eqref{eq:imp_phipm} with Eq.~\eqref{eq:imp_statA}, we
need to use the equations of motion to eliminate $(\delta
A_0^\text{stat})_1$ and insert $(\delta A_0^\text{stat})_2$. Our
identification allows us to equate the improvement coefficients at
leading order: $b_\phi^- = c_A^\text{stat} - b_A^\text{stat}$ and
$c_\phi^- = c_A^\text{stat}$. Likewise, Ref.~\cite{Kurth:2000ki}
gives the improved, renormalized vector current as
\begin{equation}
  (V^\text{stat}_R)_0 = Z_V^\text{stat}\left(1+b_V^\text{stat}\right)
  \left(
    V_0^\text{stat}
    + ac_V^\text{stat} \bar\psi \overset{\leftarrow}{D}_j \gamma_j Q
  \right),
\end{equation}
and by again using the equations of motion we obtain at leading order
$b_\phi^+ = c_V^\text{stat} + b_V^\text{stat}$ and $c_\phi^+ =
c_V^\text{stat}$.  For a chiral action, our identification leads to
$Z_A^\text{stat}=Z_V^\text{stat}$, $c_A^\text{stat}=c_V^\text{stat}$,
and $b_A^\text{stat}=-b_V^\text{stat}$, consistent with
Ref.~\cite{Ishikawa:1999yi}.

\section{Comparison with whole-operator approach}
\label{app:whole_op}

In Ref.~\cite{Chen:2017mie}, the symmetry properties of
dimension-three operators of type $\cO_\Gamma(\xi)$ and similar
dimension-four nonlocal operators were studied in order to understand
mixing and $O(a)$ effects. We find the same pattern of mixing with
three-dimensional operators and four-dimensional operators
proportional to $m_q$. However, in general it is much less
constraining to consider the operator as a whole rather than using the
auxiliary field approach to represent it using two local
operators. This leads to the following differences:
\begin{enumerate}
\item The auxiliary field approach implies that for a chiral fermion
  action, the renormalization of $\cO_\Gamma(\xi)$ is independent of
  $\Gamma$ and depends only on two parameters $Z_\phi$ and $m$. When
  chiral symmetry is broken, the splitting between different $\Gamma$
  and the mixing are controlled by a single parameter,
  $r_\text{mix}$. In contrast, the whole-operator approach implies a
  generic $\xi$-dependent and $\Gamma$-dependent
  renormalization\footnote{Note that whole-operator nonperturbative
    renormalization done using RI-MOM type
    schemes~\cite{Alexandrou:2017huk, Chen:2017mzz,
      Alexandrou:2019lfo, Shanahan:2019zcq} has generically found a
    dependence on $\Gamma$. However, this is by construction in the
    definition of observables used for imposing renormalization
    conditions. Once converted to a minimal scheme such as $\MSbar$,
    the pattern predicted by the auxiliary field approach is
    recovered~\cite{Constantinou:2017sej}, up to the precision of the
    scheme conversion.}.
\item In the auxiliary field approach, the four-dimensional operators
  with derivatives can only have those derivatives inserted in either
  local operator, i.e.\ effectively at either end of the Wilson
  line\footnote{Insertions in the Wilson line can occur as a result of
    higher-dimensional corrections to the action of the auxiliary
    field.}. By using the equations of motion for the fermion field
  and the auxiliary field, the number of improvement terms with
  derivatives can be significantly reduced. On the other hand,
  Ref.~\cite{Chen:2017mie} found operators with $\slashed{D}_\perp$ or
  $\slashed{n}(n\cdot D)$ inserted at any point $\xi'\in[0,\xi]$ along
  the Wilson line. This large number of operators makes it appear
  impractical to attempt $O(a)$ improvement.
\item The local operator $\phi=\bar\zeta\psi$ transforms under the
  fundamental irrep of the fermion flavor symmetry group. This means
  that there is no mixing of $\cO_\Gamma$ with nonlocal gluonic
  operators of the type discussed in Ref.~\cite{Zhang:2018diq}. On the
  other hand, the whole-operator approach would predict that those
  gluonic operators could have a divergent $O(a^{-1})$ contribution
  from a flavor singlet nonlocal quark operator $\cO_\Gamma$.
\end{enumerate}

\bibliography{auxiliary_field.bib}

\end{document}